\documentclass[letterpaper,twocolumn,10pt]{article}
\usepackage{usenix}

\usepackage{tikz}
\usepackage{amsmath}
\usepackage{color}

\newcommand*{\Mname}{GMI-DRL}
\usepackage{ctable} 
\usepackage{filecontents}
\newcommand\hlp[1]{\underline{\textbf{\textit{{#1}}}}}
\newcommand\todo[1]{\textcolor{red}{#1}}
\newcommand\fix[1]{\textcolor{blue}{#1}}
\usepackage{setspace}
\usepackage{listings}
\usepackage{xcolor}
\usepackage{subfig}
\usepackage{color}
\definecolor{codegreen}{rgb}{0,0.6,0}
\definecolor{codegray}{rgb}{0.5,0.5,0.5}
\definecolor{codepurple}{rgb}{0.58,0,0.82}
\definecolor{backcolour}{rgb}{0.95,0.95,0.92}
\definecolor{textblue}{rgb}{.2,.2,.7}
\definecolor{textred}{rgb}{0.54,0,0}
\definecolor{textgreen}{rgb}{0,0.43,0}
\definecolor{codered}{rgb}{201,72,12}
\usepackage[T1]{fontenc}
\usepackage[scaled=0.85]{beramono} 
\usepackage[ruled,vlined,linesnumbered]{algorithm2e}

\SetCommentSty{mycommfont}
\usepackage{multirow}
\usepackage{amsfonts}

\newcommand*\circled[1]{\tikz[baseline=(char.base)]{
            \node[circle,fill=.,inner sep=0.8pt] (char) {\textcolor{white}{#1}};}}

\begin{document}

\date{}

\title{GMI-DRL: Empowering Multi-GPU Deep Reinforcement Learning with \\ GPU Spatial Multiplexing}

\author{
{\rm Yuke Wang, Boyuan Feng, Zheng Wang, $^\dagger$Tong Geng, $^\dagger$Ang Li, and Yufei Ding}\\
\{yuke\_wang, boyuan, zheng\_wang, yufeiding\}@cs.ucsb.edu\\
University of California, Santa Barbara\\
$^\dagger$\{tong.geng, ang.li\}@pnnl.gov\\
Pacific Northwest National Laboratory
} 

\maketitle

\begin{abstract}
With the increasing popularity of robotics in industrial control and autonomous driving, deep reinforcement learning (DRL) raises the attention of various fields.
However, DRL computation on the modern powerful GPU platform is still inefficient due to its heterogeneous workloads and interleaved execution paradigm.
To this end, we propose \textbf{GMI-DRL}, a systematic design to accelerate multi-GPU DRL via GPU spatial multiplexing. 
{We introduce a novel design of resource-adjustable GPU multiplexing instances (GMIs) to match the actual needs of DRL tasks, an adaptive GMI management strategy to simultaneously achieve high GPU utilization and computation throughput, and a highly efficient inter-GMI communication support to meet the demands of various DRL communication patterns.}
Comprehensive experiments reveal that GMI-DRL outperforms state-of-the-art NVIDIA Isaac Gym with NCCL (up to 2.81$\times$) and Horovod (up to 2.34$\times$) support in training throughput on the latest DGX-A100 platform. 
%
%
Our work provides an initial user experience with GPU spatial multiplexing in processing heterogeneous workloads with a mixture of computation and communication. 
\end{abstract}

\section{Introduction}
In recent years, deep reinforcement learning (DRL) has raised significant interest of research and industry fields.
DRL combines the conventional reinforcement learning algorithms~\cite{ppo, a3c} with deep neural networks and has demonstrated superhuman performance in decision making.
DRL has been applied in many real-world applications across various domains, such as robotics~\cite{tai2017virtual,gu2017deep, zhang2015towards,kalashnikov2018scalable}, industry control~\cite{diao2019autonomous,chen2021powernet,spielberg2017deep, zhang2019deep}, and autonomous driving~\cite{sallab2017deep,wang2018deep,kiran2021deep,chen2019model }. 
\begin{figure} [t] \small
    \centering
    \subfloat[]{\includegraphics[width=\columnwidth]{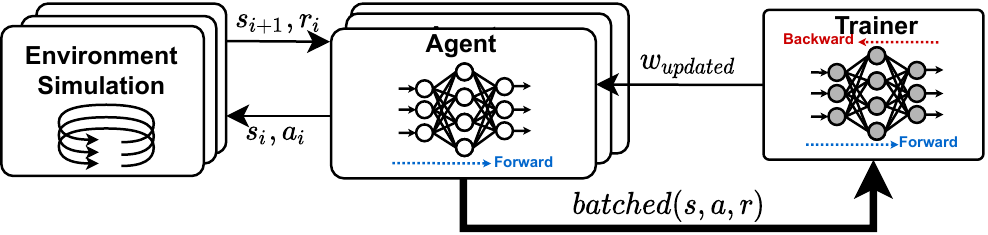}}
    \\
    \vspace{-10pt}
    \subfloat[]{\includegraphics[width=\columnwidth, height=2.6cm]{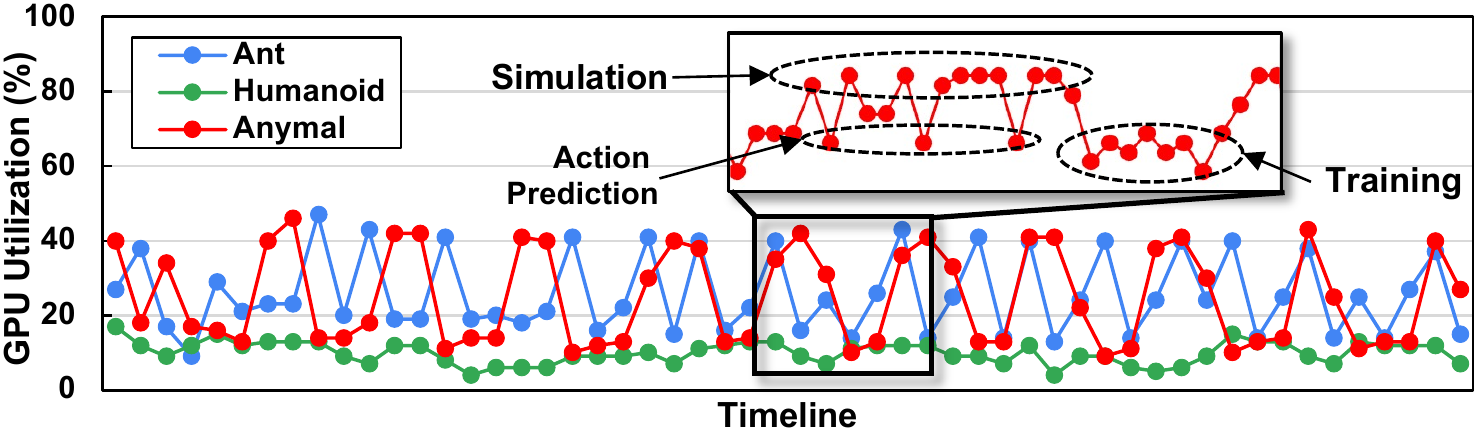}}
    \vspace{-10pt}
    \caption{(a) Basic DRL Computation Flow. $\underline{s}$: the environment state vector; $\underline{a}$: the action choice of agent; $\underline{r}$: the reward value; $\underline{w}$: weight parameters of the policy model; (b) GPU utilization of Isaac Gym for PPO training on one A100 GPU.}
    \label{fig: Performance profiling of isaac-gym on GPU utilization during the training.}
    \vspace{-10pt}
\end{figure}
\begin{figure*} [t] \small
    \centering
    \includegraphics[width=\textwidth, height=3.5cm]{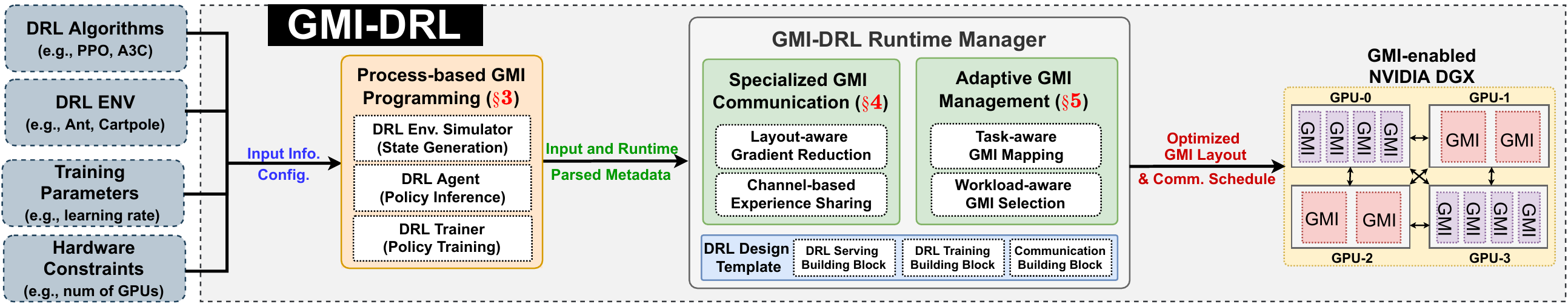}
    \vspace{-15pt}
    \caption{Overview of \Mname.}
    \label{fig: Overview of GMI-DRL.}
    \vspace{-15pt}
\end{figure*}

{
DRL features heterogeneous computation workloads and multi-granularity communication patterns (Figure~\ref{fig: Performance profiling of isaac-gym on GPU utilization during the training.}(a)). It interleaves the execution of three components: \textit{Environment Simulator}, \textit{Agent}, and \textit{Trainer}. 
Agents will interact with environment simulators through fine-grained state-action sharing to collect enough experience (consisting of state, action, and reward). 
After a number of (e.g., 32 and 64) agent-environment interactions, the agent will communicate with a trainer via coarse-grained data sharing for a batch of experience.
The above three components feature different computation patterns. 
The environment simulator relies on physics simulation (e.g., local motion and robotics control) while the agent and trainer depend on GEMM-based neural network operations (e.g., policy model inference/training). 
} 

Despite the stunning success of DRL, DRL execution is still far from efficient on modern computing platforms, such as GPUs. 
For illustration, we choose the state-of-the-art NVIDIA Isaac Gym~\cite{isaac-gym} (single-GPU DRL simulation with hundreds/thousands of environments and agents running in parallel) on PPO~\cite{ppo} algorithm. Figure~\ref{fig: Performance profiling of isaac-gym on GPU utilization during the training.}(b) gives the profiling results of GPU utilization on three representative DRL benchmarks for 10 training epochs. 
%
We observe that the overall GPU utilization is consistently under 50\% (32\% on average). 
Such low GPU utilization mainly comes from the poor scalability of environment simulation (sophisticated physics calculation) and its interleaved execution with other tasks (action prediction and policy training) with different computation paradigms. 
In addition, with more complex real-world DRL applications, a huge amount of training samples (experience) are thus in demand for learning complex DRL tasks quickly. This drives the need for scaling up both DRL simulation and policy training on powerful multi-GPU platforms (e.g., NVIDIA DGX~\cite{dgx-a100}).
Unfortunately, existing work~\cite{cule, a3c, isaac-gym, liang2018gpu} of DRL either only focuses on single-GPU setting or naively scaling up single-GPU designs to multi-GPU platforms without dedicated distributed designs and optimizations to maximize the GPU utilization and minimize communication overhead. Thus, it could hardly exploit the full potential of the powerful multi-GPU system.

Our key insight is that DRL performance can potentially be improved by turning those ``spare'' GPU resources into full utilization.
A promising direction is to harness \textit{GPU spatial multiplexing} (e.g., MPS~\cite{nvidia-mps} and MIG~\cite{mig}). It opens a new way for fine-grained utilization of GPU resources, where a large GPU can be used as a set of small sub-GPUs for different tasks. This also enlarges the existing design and optimization space for using individual GPUs. 
For instance, we can treat one GPU as a multi-sub-GPU system for computation and communication at the sub-GPU level. 
When scaling up to multiple GPUs, GPU spatial multiplexing is also an effective way for better resource management of different workloads. It can facilitate the creation of the ``best fit'' GPU resource plans, given the specified number of GPUs and the GPU affinity demand (e.g., GPUs with high-speed inter-GPU connections).

However, GPU spatial multiplexing is not low-hanging fruit. 
GPU spatial multiplexing introduces the ``memory barrier'' (i.e., memory isolation) among different sub-GPUs. 
This would be the performance killer for DRL with heterogeneous tasks and lots of inter-task communication. 
Efficient communication at the sub-GPU level is still missing from existing high-performance GPU communication libraries (e.g., NCCL~\cite{nccl}). 
Such communication hurdles may offset computation gains of GPU spatial multiplexing.
Besides, finding the right-fit sub-GPUs is non-trivial. 
The mismatch between sub-GPUs and their assigned DRL workloads would easily downgrade the overall performance while wasting valuable GPU resources (e.g., SM and memory). 
Furthermore, identifying the optimal layout of sub-GPUs requires careful consideration of the specialties of workloads (e.g., communication demands) and hardware (e.g., GPU interconnections). 

To this end, we introduce \Mname, a systematic design for accelerating multi-GPU deep reinforcement learning via GPU spatial multiplexing (Figure~\ref{fig: Overview of GMI-DRL.}). 
\Mname~addresses the key inefficiency of DRL computation (e.g., low GPU utilization) and effectively scales up to multi-GPU platforms for handling many real-world large-scale DRL applications (e.g., power-grid control~\cite{chen2021powernet}). 
%
\Mname~incorporates a new concept -- \textit{GPU Multiplexing Instance} (GMI), a unified resource-adjustable sub-GPU design for heterogeneous tasks.
%
%
In short, \Mname~has several highlights:

\textit{\textbf{1) \Mname~is communication-effective.}} 
It covers various types of communication patterns across GMIs, including the latency-optimized cross-GMI gradient reduction for synchronized DRL training and throughput-optimized cross-GMI experience sharing for asynchronized DRL training. 

\textit{\textbf{2) \Mname~is computation-adaptive.}} It features an adaptive management of GMIs, which encompasses several GMI layout templates to maximize the computation and communication efficiency in DRL and a new heuristic approach to determine the resource budget (e.g., SMs and device memory) of GMIs to fit heterogeneous DRL workloads.

{\textit{\textbf{3) \Mname~is workload- and hardware-aware.}} 
It offers user-friendly APIs for capturing key characteristics of DRL workloads to guide our runtime configuration search. 
It also offers MPS~\cite{nvidia-mps} and MIG~\cite{mig} as the backend and can automatically determine which one to use according to the actual needs (e.g., communication) of DRL applications and hardware specifications (e.g., GPU architecture).}  

\Mname~is the first work to explore the great potential of GPU spatial multiplexing by combining the advantages of MPS and MIG. More importantly, \Mname~encapsulates generic programming, computation, and communication paradigm to cover not only DRL but also a wide range of applications (e.g., Gene Alignment~\cite{morgenstern1996multiple} and Stencil~\cite{ragan2013halide}). This will facilitate resource-efficient and high-performance computing on GPUs in the future. 

To conclude, we summarize our contributions as
\begin{itemize}
    \vspace{-5pt}
    \item We identify the key performance bottleneck of GPU-based DRL and explore the great potential of GPU spatial multiplexing for accelerating DRL. 
    \vspace{-5pt}
    \item We introduce a new notion of GPU multiplexing instance (GMI) both physically (at the hardware-resource level) and logically (at the programming-model level). This enables the utilization of GPUs in a more fine-grained and resource-efficient manner and enlarges the space for designs and optimizations ($\S$\ref{sect: overview of GMI-DRL}).
    \vspace{-5pt}
    \item We design a highly efficient and generic data communication layer for GMIs (cross-GMI collective and peer-to-peer communication) to facilitate DRL training with diverse communication demands ($\S$\ref{sect: specialized GMI communication}). 
    \vspace{-5pt}
    \item  We propose an adaptive GMI management strategy, which encompasses task-aware GMI mapping and workload-aware GMI selection technique, for better computation performance and GPU utilization. ($\S$\ref{sect: Adaptive GMI Management})
    \vspace{-5pt}
    \item Extensive experiments show that \Mname~outperforms the state-of-the-art NVIDIA Isaac Gym with NCCL and Horovod as the backend in computation throughput on NVIDIA DGX across various DRL benchmarks.
\end{itemize}

\section{Background}
\label{sect: Background}
In this section, we will introduce the basics of the existing GPU spatial multiplexing techniques.

\textbf{Multi-processing Service (MPS):}~\cite{nvidia-mps} is an alternative, binary-compatible implementation of the CUDA Application Programming Interface (API), which can transparently enable co-operative multi-process CUDA applications.
There are three major components of the MPS: 
1) \textit{Control Daemon Process: }The control daemon manages the start and stop of the MPS server and coordinates connections between MPS clients and MPS servers.
2) \textit{Client Runtime: }The MPS client runtime is built into the CUDA driver library and may be used transparently by any CUDA application. Note that each process created by the user-defined application would be handled by one MPS client when it intends to use GPUs.
3) \textit{Server Process:} The server is the clients' shared connection to the GPU and provides concurrency among clients.
{In order for applications to benefit from MPS, several basic requirements must be satisfied: (i) the application should execute in the fashion of multi-process parallelism, and  
(ii) the individual process should underutilize the GPU resources. 
}  

\begin{figure} [t] \small
    \centering
    \includegraphics[width=\columnwidth]{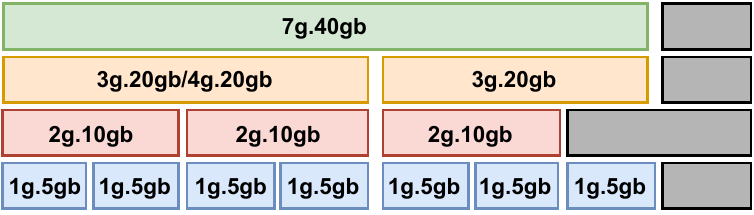}
    \vspace{-10pt}
    \caption{Available MIG instance combination on A100. Note that grey boxes indicate reserved compute/memory space. ``2g.10gb'' represents a MIG instance with $\frac{2}{8}$ of the total number of GPU SMs and 10 GB device memory.}
    \label{fig: GI Combination}
\end{figure}

\textbf{Multi-Instance GPU (MIG)}~\cite{mig} {is a new GPU spatial multiplexing solution introduced in NVIDIA Ampere GPUs (e.g., A100), which can offer higher hardware-resource-level isolation compared to MPS.}
The main design goal of MIG is to physically divide a large GPU for different users running \textit{multiple independent jobs/applications} (e.g., data processing, NN inference, and NN training) without any interaction among each other.
MIG offers diverse combinations of GPU instances to utilize GPUs in a resource-efficient way. As exemplified in Figure~\ref{fig: GI Combination}, the computation resources of an A100 GPU are divided into 8 partitions (7 partitions usable while 1 partition is reserved). 
%
This would help to easily adapt the GPU for different application purposes, such as co-locating NN training and inference on the same GPU. 

\begin{table}[t] \small
\centering
\caption{Comparison between MIG and MPS. Note that ``Iso.'': Isolation; ``QoS'': Quality of Service; ``Com'': Communication among different instances.}
\vspace{-5pt}
\scalebox{0.8}{
\begin{tabular}{l|c c c c c}
\specialrule{.1em}{.05em}{.05em} 
\textbf{Technique} & \textbf{Partition} & \textbf{Mem. QoS} & \textbf{SM Iso.} & \textbf{Error Iso.} & \textbf{Com.} \\ \hline
MPS         & Logical    & No       & By Percentage & No  & Yes\\ 
MIG         & Physical   & Yes      & By Partition  & Yes & No\\ 
\specialrule{.1em}{.05em}{.05em} 
\end{tabular}}
\label{tbl: qualtitive comparison of existing techniques}
\vspace{-15pt}
\end{table}

Table~\ref{tbl: qualtitive comparison of existing techniques} gives a qualitative analysis of the difference between MPS and MIG.
The existing work on MIG and MPS mainly focuses on the settings with many independent jobs/applications, such as DNN model inference/serving~\cite{tan2021serving, choi2021multi}. 
Our work is the first to exploit the potential of GPU spatial multiplexing techniques to improve the performance of a \textit{single job/application} (e.g., DRL training) with \textit{heterogeneous tasks} (e.g., simulation, action prediction, and policy training) that closely interact with each other.
We remark there are several key questions to be addressed:
1) \textit{How to effectively share data between GPU instances so that the communication overhead can be minimized;}
2) \textit{How to effectively partition the GPU resources so that the demands of heterogeneous tasks of an application/job (\textit{e.g.}, the latency of policy inference and throughput of policy training) can be fulfilled and the overall system performance can be maximized.}
\section{Overview of \Mname}
\label{sect: overview of GMI-DRL}
In this section, we will discuss the overview of \Mname~from the end-user perspective.

\textbf{Process-based GMI Programming} 
\Mname~is built with a process-based programming paradigm for better service-level performance guarantee and less computation interference~\cite{mig, nvidia-mps}. 
Different processes would have configurations (e.g., environmental variables) and memory address space that are independent of each other. 
To facilitate the programming of the DRL, we introduce the base class, \texttt{DRL\_role}, which consists of several major functions, as shown in Listing~\ref{code: basic definition of a RL role}. 
The initialization function will set up the GMI based on its \texttt{GMI\_id} and register a GMI to a global GMI manager. 
It will also attach its GMI to a specific GPU based on the provided GPU id. 
Once it has been registered, the GMI will also be assigned to a GMI group for inter-GMI communication if necessary. 
There are several other functions, including the major GMI execution routine (\texttt{GMI\_run}) and communication primitives (e.g., \texttt{collective} and \texttt{p2p}), to be implemented depending on the specific purpose of the DRL\_role (e.g., simulator and trainer). {We determine the GMI backend (MIG or MPS) based on the actual need of DRL tasks and GPU architectures. On the DRL training, we choose MPS for better communication efficiency while for DRL serving, we select MIG for better computation performance. On GPUs with 70$\leq$$sm$$<$80 (e.g., V100), only MPS is available. While for GPUs with $\mathit{sm}\text{=}$80 (e.g., A100), MPS and MIG are available.
We detail more comparison of GMI backends at $\S$\ref{sect: additional studies}.}
\begin{figure}[t]
\vspace{-12pt}
\begin{lstlisting}[caption={Example of GMI-based Programming.}, label={code: basic definition of a RL role}]
import GMI_DRL 
# import other packages ...
GMI_DRL.config.num_GPUs=3        # GPU configurations.
GMI_DRL.config.GPU_arch="sm_80"  # sm_80(A100),sm_70(V100)
GMI_DRL.config.enable_NCCL=True  # NCCL availability.
GMI_DRL.config.GMI_backend="MPS" # GMI backends.
# define class for DRL tasks.
class DRL_role(object):
    # Initialize the base environment.
    def __init__(self, GMI_id, role, gpu_id):
        self.GMI_id = GMI_id
        self.role = role
        self.mgr = GMI_DRL.GMI_manager.add_GMI(GMI_id)
        self.mgr.set_GPU(GMI_id, gpu_id)
        self.group = self.mgr.get_group(GMI_id, GMI_id)
    def GMI_run(self, param1, param2, ...):
        # major routine of DRL tasks,
        # e.g., simulators, agents, and trainers.
    def GMI_collective(self, data):
        # some data processing work ...
        proc_data = proc_fun(data)
        # allreduce data within a group of GMIs. 
        self.mgr.allreduce(proc_data, self.group)
    def GMI_send(self, data, dst_GMI_id):
        # some data processing work ...
        proc_data = proc_fun(data)
        # Asynchronously send data to another GMI.
        self.mgr.send(proc_data, dst_GMI_id)
    def GMI_recv(self, src_GMI_id):
        # Synchronously receive data from another GMI.
        data = self.mgr.recv(src_GMI_id)
\end{lstlisting} 
\vspace{-20pt}
\end{figure}

\textbf{DRL Environment Simulator} is the most essential component for generating environment states and action rewards. 
Each simulation can hold hundreds or thousands of environments running concurrently that will interact with agents. 
To create a DRL simulator in GMI, users should create and initialize the simulation engine at the \texttt{init} function. 
Simulation configurations, such as the number of environments, should also be specified at the time of simulation engine creation. 
The \texttt{run} function of a simulator will be a loop with a specified upperbound (e.g., the total number of training iterations). 
In each serving or training iteration, simulators will interact with agents by receiving the vectors of actions and old states from the agent and then sending new states and rewards back to the agent. 
In general, due to performance consideration, we colocate the environment simulator with the agent on the same GMI for ease of data sharing (detailed in $\S$\ref{sect: Adaptive GMI Management}). Thus, inter-GMI communication primitives (e.g., collective communication) will not be applied to the DRL simulator.

\textbf{DRL Agent} is the key component to make action decisions based on the environment states. 
Similar to the implementation of the simulator, users will implement a \texttt{GMI\_run} function to interact with a simulator in an iterative manner. 
Depending on the needs of DRL applications, agents would be responsible for communicating with other GMIs in two different ways. 
In the synchronized training settings (e.g., PPO~\cite{ppo} algorithm), agent and trainer are co-located in the same GMI ($\S$\ref{sect: Resource-aware GMI Mapping}), agents will share the collected experience vectors (consisting of state, action, and reward) with a trainer and then wait for the completion of the model update. 
In the asynchronized DRL training, the experience collected by agents will be shared with trainers on different GMIs. Communication function \texttt{send\_p2p} will be implemented for experience movements and policy model synchronization.

\textbf{DRL Trainer } is the major component for updating the policy model based on the collected experience data from agents. The \texttt{run} function will be the loop iteration to receive the experience and update models iteratively. 
During the training, the trainer will feed the experience to the policy NN model for the forward and backward computation to generate the policy gradients.
%
Next, trainers will rely on the collective communication primitives (e.g., AllReduce) for synchronizing policy gradients with other trainers to maintain a consistent policy model. 
Finally, each trainer will use the synchronized policy gradients to update the policy model.
Depending on the underlying GMI mapping of DRL trainers (on the same GPU or across GPUs), different options will be automatically determined for efficient policy model synchronization ($\S$\ref{sect: Layout-aware Gradient Reduction}).
\section{Specialized GMI Communication}
\label{sect: specialized GMI communication}
DRL is a communication-intensive application that requires frequent information exchange among different components. 
However, GMI-based design introduces memory barriers that complicate communication.
To this end, we introduce a \textit{specialized GMI communication} design to ease the communication efforts.
We tackle two major types of communication in DRL training: 1) latency-optimized policy gradient reduction for synchronized DRL training (e.g., PPO~\cite{ppo}) among different trainers and 2) throughput-optimized experience sharing in the asynchronized DRL training (e.g., A3C~\cite{a3c}) between agents and trainers. 
Communication between simulator and agent is omitted since it is simple and happens within the same GMI due to performance consideration ($\S$\ref{sect: Adaptive GMI Management}). Policy parameter sharing between agents and trainers in asynchronized training is skipped due to its very minor performance impact ($<$5\%) compared to experience sharing based on studies. 
\begin{figure*} [t] \small
    \centering
    \includegraphics[width=2\columnwidth]{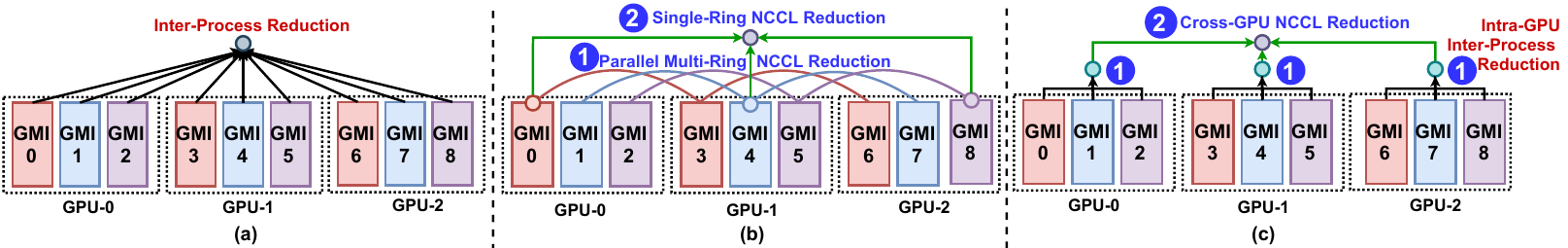}
    \vspace{-5pt}
    \caption{Exploration of Inter-GMI Policy Gradient Reduction: (a) Multi-Process Reduction; (b) Multi-Ring Reduction; (c) Hierarchical Reduction. Note that the final step for broadcasting the result to all GMIs is omitted for simplicity.}
    \label{fig: Exploration of Inter-GMI Communication}
    \vspace{-5pt}
\end{figure*}

\vspace{-5pt}
\subsection{Layout-aware Gradient Reduction} 
\label{sect: Layout-aware Gradient Reduction}
During the synchronized DRL training, to maintain a consistent view of the policy model, we need an efficient way of communication with low latency overhead.
In the data-parallel DNN training on multi-GPU platforms, a set of highly optimized communication primitives (e.g., NCCL~\cite{nccl}, NVIDIA IPC~\cite{nvipc}) have been developed.
However, these primitives are tailored for inter-GPU data movements at the granularity of GPUs (one process per GPU) and \textbf{cannot} be directly applied to inter-GMIs communication at the granularity of sub-GPUs (multiple processes per GPU). 
This is because existing GMI backend designs, such as MPS~\cite{nvidia-mps} and MIG~\cite{mig}, are enforcing inter-process memory isolation to avoid any memory security issues, such as side-channel attacks~\cite{nvidia-mps, mig}. 
Therefore, data (e.g., tensors) could not be shared among GMIs even if they are concurrently sharing the same physical GPU. To this end, we introduce a novel \textit{layout-aware gradient reduction} (LGR) strategy, including three potential solutions:

\textbf{Multi-Process Reduction (MPR): } This solution treats the inter-GMI communication as the pure inter-process synchronization problem, as shown in Figure~\ref{fig: Exploration of Inter-GMI Communication}(a).
It requires first to move the policy gradient from GMI to CPU and then apply gradient reduction on the CPU across different processes.
This is the basic solution for handling gradient reduction among GMIs, and it is the most generic solution to handle any GMI layout regardless of their underlying placement on GPUs, such as on the same GPU or across multiple GPUs. 
%
However, in the more complex GMI layouts (e.g., GMIs across multiple GPUs), this strategy comes with several weaknesses:
1) incurring an excessive amount of data traffic between CPU host and GPU device memory; 
2) idling the high-speed inter-GPU hardware infrastructures, such as NVLink and NVSwitch;
3) relying on the slow CPU for reduction computation. 

\textbf{Multi-Ring Reduction (MRR): }
For GMI layout with cross-GPU reduction demands, counting on the existing inter-GPU communication infrastructure, such as NCCL for ring-based reduction via NVLink, would be a promising option. 
Note that while GMI prevents the synchronization at the same GPU via NCCL primitives, it allows the synchronization among multiple GMIs (each placed on different GPUs) to communicate via NCCL.
Figure~\ref{fig: Exploration of Inter-GMI Communication}(b) shows that GMIs on different GPUs are forming multiple non-intersect NCCL rings for NVLink communication, as illustrated as the red, blue, and purple curves at Step \circled{1}.
Once all rings' reduction complete, we initiate another NCCL ring reduction to synchronize all partial results from the prior stage of multi-ring reduction, illustrated as the green lines at Step \circled{2}. 
While this strategy can leverage the high-speed inter-GPU communication, it is limited to the setting when the number GMI per GPU is less or equal to the number of GPUs. 
Otherwise, the last synchronization NCCL ring has to cover more than one GMI (i.e., the endpoints of more than one ring) on the same GPU, which will trigger ``multiple CUDA streams error'' in NCCL.
\begin{algorithm}[t] \footnotesize
\setstretch{0.96} 
  \caption{Communication Strategy Selection.}
  \label{algo: Communication Strategy Selection.}
\SetAlgoLined
  \SetKwInOut{Input}{input}
  \SetKwInOut{Output}{output}
  \Input{GMI-to-GPU Mapping List $\mathit{MPL}=[[0,1,2],[3,4,5],[6,7,8],..,]$.}
  \Output{Selected Communication Strategy.}
    $\mathit{GMIperGPU\_set = set()}$\;
    \tcc{When all GMIs on the same GPU}
    \If{len($\mathit{MPL}$) $\leq$ $1$}{
        \textbf{return} ``Multi-Process Reduction''\;
    }
    \tcc{Traverse all GMIs on all GPUs.}
    \For{$\mathit{GMI\_li}$ \textbf{in} $\mathit{MPL}$}{
        $\mathit{GMIperGPU\_set.add(len(GMI\_li))}$;
    }
    \tcc{Different GPUs have different number of GMIs}
    \If{$\mathit{GMIperGPU\_set.size() > }1$}{
        \textbf{return} ``Hierarchical Reduction''\;
    }
    \tcc{The number of GMIs per GPU is greater than the total number of GPUs}
    \If{$\mathit{GMIperGPU\_set.pop() > len(MPL)}$}{
        \textbf{return} ``Hierarchical Reduction''\;
    }
    \textbf{return} ``Multi-Ring Reduction''\;
\end{algorithm}
\setlength{\textfloatsep}{1pt}

\begin{figure*} [t] \small
    \centering
    \includegraphics[width=\textwidth]{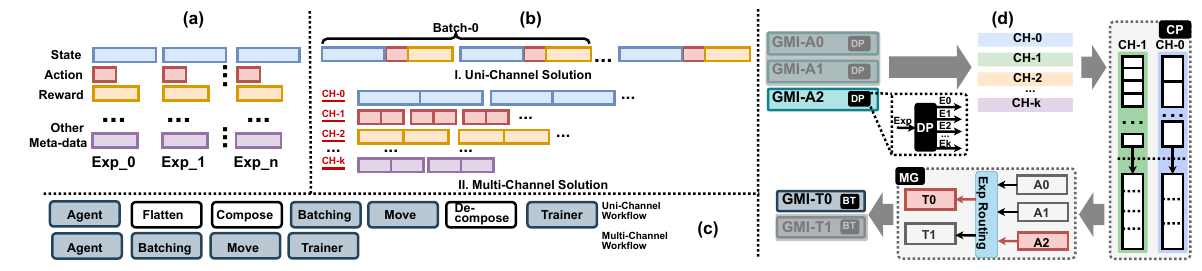}
    \vspace{-15pt}
    \caption{Multi-Channel Communication. (a) Major components of experience (``Exp\_*'') for DRL training; (b) Comparing data communication patterns in the uni-/multi-channel design; (c) Comparing data communication workflow in the uni-/multi-channel design; (d) Exemplification of Channel-based Experience Sharing (from A2 to T0). Note that ``GMI-A*'' is the GMI for DRL agents while ``GMI-T*'' is the GMI for DRL trainers; ``E0...Ek'' are categorized experience data for different channels (CH).}
    \label{fig: Illustration of Multi-Channel Communication.}
    \vspace{-10pt}
\end{figure*}

\textbf{Hierarchical Reduction (HAR):}
To balance the adaptability and performance of previous two solutions, we introduce a \textit{hierarchical reduction} (HAR) technique to efficiently handle the synchronized DRL training based on GMIs (Figure~\ref{fig: Exploration of Inter-GMI Communication}(c)).  
The key idea is to systematically combine different types of communication strategies (e.g., Multi-Process and Multi-Ring) at different levels (e.g., inter-GMI or inter-GPU) of communication to (i) maximize the communication parallelism, (ii) minimize unnecessary data movement, and (iii) exploit the benefits of powerful hardware interconnects and communication libraries. 
HGR design consists of two major steps: 
Step \circled{1} is to synchronize policy model gradients among trainer GMIs on the same GPU. 
%
Step \circled{2} is cross-GPU gradient synchronization, which will carry out AllReduce operation among GMIs belonging to different GPUs. 
The key of this step is to select one GMI (``leader'' GMI) from each GPU that will be involved in this inter-GPU synchronization.
Here we identify the leader GMI if its ID satisfies $\mathit{GMI\_id\%M = t}$, where $0\leq t < M$ and $M$ is the number of GMIs per GPU.

In practice, we determine the choice of the aforementioned three options based on the DRL trainer GMI layout, as illustrated in Algorithm~\ref{algo: Communication Strategy Selection.}. We take the GMIs and their located GPUs at the input. For example, the input list $\mathit{MPL}=[[0,1,2],[3,4,5],[6,7,8],..,]$ can be interpreted as GMIs with ID$=\{0,1,2\}$ are located on GPU=0.  
We further give a theoretical time complexity comparison among these three methods, as shown in Table~\ref{tbl: time complexity comparison}, where $t$ is the number of GMIs per GPU; $g$ is the number of GPUs; $B_1$ is the inter-GMI bandwidth via inter-process communication;
$M_p$ is the size of policy model parameters; $B_2$ is the bandwidth via NCCL. 

After the reduction through either of the above three strategies, we will flush back the synchronized gradients towards all GMIs through gradient broadcasting. Compared with the prior steps of policy gradient synchronization (interleaved with communication and computation), the gradient broadcasting primitive is low-cost, since it can be easily parallelized without operation dependency. 
\begin{table}[t] \small
\centering
\caption{Time complexities of MPR, MRR, and HAR.}
\vspace{-5pt}
\scalebox{1}{
\begin{tabular}{l|c}
\specialrule{.1em}{.05em}{.05em} 
\textbf{Solution} & \textbf{Time Complexity} \\ \hline\hline
MPR & $2 \cdot (g\cdot t - 1) \cdot M_p/(g\cdot t\cdot B_1)$ \\ \hline
MRR & $2 \cdot (g - 1) \cdot (t+1)\cdot M_p/(g\cdot B_2)$\\ \hline
HAR & $2\cdot (g - 1)\cdot M_p/(g\cdot B_2) + 2\cdot (t - 1) \cdot M_p/(t\cdot B_1)$ \\
\specialrule{.1em}{.05em}{.05em} 
\end{tabular}}
\label{tbl: time complexity comparison}
\vspace{5pt}
\end{table}
%

\vspace{-5pt}
\subsection{Channel-based Experience Sharing} 
\label{sect: Channel-based Experience Sharing}
As the key components for asynchronized DRL training (e.g., A3C~\cite{a3c}), cross-GMI experience sharing is to connect the agent GMIs and trainer GMIs. 
Compared with the gradient reduction in the synchronized DRL training, the size of the experience data is bulky, which demands a throughput-optimized connection and transferring strategy to achieve higher training throughput performance.
%
%
We introduce our multi-channel solution to tackle the experience transferring bottleneck systematically. 
A brief illustration in Figure~\ref{fig: Illustration of Multi-Channel Communication.}(a)(b)(c) would demonstrate the advantages of our design conceptually, such as operation saving, batching and transferring efficiency. 
%
Our key design insight is that the experience data granularity at the collection, transmission, and training could be different and would be dedicatedly optimized for maximizing the performance at each step.
To achieve such variable transmission granularity, we incorporate several components (Figure~\ref{fig: Illustration of Multi-Channel Communication.}). 

The first component is the experience \textit{dispenser (\textbf{DP})}, which is attached to each agent as asynchronized backend service. The major task of this service is to categorize the experience data and push them into corresponding output channels, such as the state channel and reward channel. There are major reasons for creating different channels for different types of data in the experience: 
1) more flexibility in handling data collection and the data transferring at different levels of granularity and transmission rate; 2) more efficiently handling data with different sizes within the same batch of experience.

The second component is the experience \textit{compressor (\textbf{CP})}, which is a system-wide service that controls the experience sharing of each channel for all agent GMIs. 
It will manage the size of each data transmission so that the data movement among different GMIs can be optimized for achieving high throughput. One of the typical operations for CP is to concatenate multiple tensors from the same channel to increase the size of each data movement.

The third component is the experience \textit{migrator (\textbf{MG})}, which is a system-wide service that manages the routing of experience from agents to trainers. 
If the agent and trainer GMIs are located at the same GPUs. MG will directly forward the data by channels to trainers. 
If the agent and trainer GMIs are located at different GPUs. MG will first gather experience data by channels among GMIs and then distribute experience by channels to trainers with the least workload.

The fourth component is the experience \textit{batcher (\textbf{BT})}, which is attached to each trainer GMI. It conducts data preparation and model training. The key operations in data preparation include \textit{slicing} (if a small batch is required) and \textit{stacking} (if a large batch is required). 
BT can be optimized for different objectives depending on the actual needs of DRL tasks.  For instance, for complicated robotic simulation, a smaller batch is used for the higher model update frequency. While in other settings, such as local motion simulation, a larger batch could be used to reduce the impact of data noise.
\section{Adaptive GMI Management} 
\label{sect: Adaptive GMI Management}
In this section, we will detail our adaptive GMI management strategy for DRL computation, including \textit{task-aware GMI mapping} for coarse-grained mapping from DRL tasks to GMIs and \textit{workload-aware GMI selection} for fine-grained configuration adjustment of DRL tasks and GMIs to maximize the DRL execution efficiency.
\begin{table}[t] \small
\centering
\vspace{-5pt}
\caption{Terms and their Meanings.}
\vspace{-8pt}
\scalebox{0.95}{
\begin{tabular}{l|p{6.5cm}}
\specialrule{.1em}{.05em}{.05em} 
\textbf{Term} & \textbf{Meaning} \\ \hline\hline
$R_s$, $R_a$, $R_t$ & the size of dominant resources for the simulator, agent, and trainer, respectively.\\ \hline
$\mathbb{I}$ & the type of dominant resource: SM or Memory. \\ \hline
$S$, $A$, $W$ & the size of a single {state}, {action}, and {reward} vector, respectively. \\ \hline
$T_s$, $T_a$, $T_t$ & the {execution time} for the simulator, agent, and trainer in each DRL training/serving iteration. \\ \hline
$BW$ & the bandwidth of inter-GMI communication. \\ \hline
$M_p$ & the size of the policy model. \\\hline
$m$ & the number of simulation steps before a training.\\\hline
$n$ & the total number of GMIs in the current system.\\ \hline
$\alpha$, $\beta$ & the size ratio factor when multiple environment simulators are sharing agents or trainers. \\
\specialrule{.1em}{.05em}{.05em} 
\end{tabular}}
\label{tbl: term and meaning.}
\vspace{5pt}
\end{table}

\subsection{Task-aware GMI Mapping} 
\label{sect: Resource-aware GMI Mapping}
{To find the best layout of DRL tasks on GMIs, we introduce a \textit{task-aware GMI mapping} by carefully considering the specialties of various DRL tasks (e.g., computation and communication) and hardware platforms (e.g., GPUs and GPU inter-connections). We incorporate such design insights into our high-efficient design templates for DRL tasks on GMIs for DRL serving (computation-dominated) and DRL training (computation-communication-hybrid).}

\textbf{DRL Serving: } 
DRL serving aims at collecting enough experience through continuous environment-agent interaction.
It consists of two components: \textit{environment simulator} and \textit{agent}. 
Optimizing DRL serving is to achieve \textit{low latency} at each round of interaction between environment simulator and agent so that more experience can be collected quickly. 
The interaction between environment simulator and agent requires \textit{frequent and fine-grained} information exchange to achieve high serving performance.
Unfortunately, in GMI-based design, communication across GMI incurs non-trivial costs because of the memory isolation enforced by existing GPU multiplexing techniques~\cite{mig, nvidia-mps}. 
Thus, data movement among different GMIs (even if they belong to the same GPU) has to bypass CPU host memory.
%
Such a constraint makes the design option (\textit{Task-dedicated GMI}, namely TDG) with dedicated GMIs for environment simulator and agent largely inefficient due to the high overhead of sharing the state, action, and reward data consistently across GMIs boundaries.
\begin{figure} [t] \small
    \centering
    \includegraphics[width=\columnwidth]{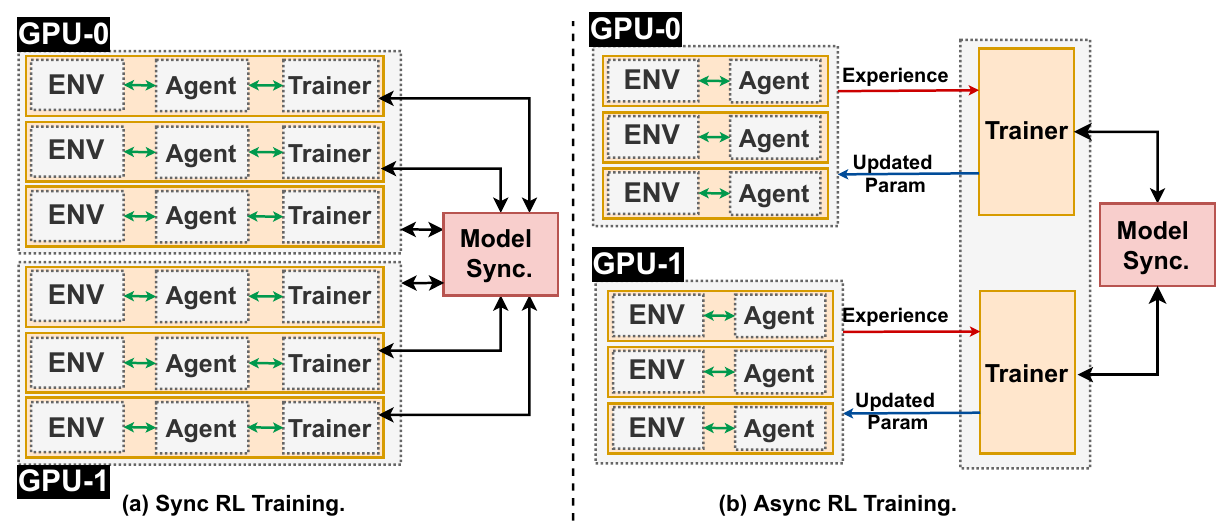}
    \caption{GMI Mapping for \textit{Sync.} and \textit{Async.} DRL Training.}
    \label{fig: GMI Mapping Template for Sync. and Async. DRL Training.}
    \vspace{5pt}
\end{figure}
Based on this observation, we propose the idea of a DRL serving block (\textit{Task-colocated GMI}, namely TCG), where each GMI manages the sequential execution of the environment simulator and agent.
In this design, the \textit{state} data generated from the environment simulator can be easily shared with the subsequent agent for action prediction via low-cost intra-GMI memory access. 
Similarly, the \textit{action} vector from agents can be easily retrieved by simulators for the next round of interaction without inter-GMI data movement ($COM=0$). 
To compare TCG with TDG, we give an analysis in terms of their resource and communication cost. We first define the terms (Table~\ref{tbl: term and meaning.}) and then detail our comparison in Table~\ref{tbl: comparison between TDG and TCG in RL serving}). 
\begin{table}[t] \small
\centering
\caption{Comparing TDG and TCG in DRL serving.}
\vspace{-5pt}
\scalebox{1}{
\begin{tabular}{l|>{\centering\arraybackslash}p{2.8825cm}|>{\centering\arraybackslash}p{2.8cm}}
\specialrule{.1em}{.05em}{.05em} 
\textbf{Solution} 
& \textbf{Resource Size} (\textbf{$R^{\mathbb{I}}$}) 
& \textbf{Comm. Size} ($COM$) 
\\ \hline\hline
TDG 
& \footnotesize{$(T_s\cdot R_s + T_a\cdot\alpha\cdot R_a)/(T_s+T_a)$} 
& \footnotesize{$2\cdot S + A + W$} 
\\ \hline
TCG 
& \footnotesize{$(T_s + T_a)\cdot\max\{R_s, R_a\}/{(T_s+T_a)}$} 
& 0  \\ 
\specialrule{.1em}{.05em}{.05em} 
\end{tabular}}
\label{tbl: comparison between TDG and TCG in RL serving}
\vspace{5pt}
\end{table}
Firstly, to determine $\mathbb{I}$ for $R_s$, $R_a$, and $R_t$, we use 
\begin{equation} \label{eq: Dimension-worker selection} \small
     \mathbb{I} = 
    \begin{cases}
     SM               &   \frac{R_{SM}}{R_{\mathit{SMperGPU}}} \geq \frac{R_{\mathit{Mem}}}{R_{\mathit{MemPerGPU}}} \\ 
     \mathit{Memory}     &   \mathit{Otherwise}.
    \end{cases}
\end{equation} 
where $R_{SM}$ and $R_{Mem}$ are computing and memory resource consumption measured on a single DRL serving/training process running exclusively on a GPU. 
In most cases, $R_{SM}$ is usually the dominant factor.
Our empirical studies show that $\alpha$ is around 0.2 and $R_s \approx 10R_a$, $T_s \approx 6T_a$, thus, the overall serving throughput can be calculated as
\begin{equation} \small
    \mathit{TOP}_{\mathit{serving}} = \frac{R^{\mathbb{I}}_{all}}{R^{\mathbb{I}}}\cdot\frac{1}{T_s+T_a+\frac{COM}{BW}}
\end{equation}
Considering Table~\ref{tbl: comparison between TDG and TCG in RL serving} and $\frac{COM}{BW} \approx 2\cdot (T_s+T_a)$ based on massive profiling, we can estimate that the overall serving throughput of our TCG solution would be higher (about $2.5\times$) compared with TDG. This is because the resource penalty (i.e., decreasing of parallelism) of simulator-agent co-locating on throughput is minor 
($(T_s + T_a)\cdot\max\{R_s, R_a\}/(T_s\cdot R_s + T_a\cdot\alpha\cdot R_a) - 1 \approx 0.16\times$)
compared with the throughput benefits (around $3\times$) by saving inter-GMI data movements. 

\textbf{Synchronized DRL Training: } 
Synchronized DRL training is the dominant paradigm (e.g., PPO~\cite{ppo}). It consists of three key components: \textit{environment simulator}, \textit{agent}, and \textit{trainer}.
There are two major stages: (i) \textit{experience collection} (multiple rounds (e.g., 32 and 64) of interaction between environment simulator and agent) and (ii) \textit{model training} (update policy model based on experience). These two stages happen sequentially at each training iteration and the training performance is sensitive to the latency of both stages.

The major goal of optimizing synchronized DRL training is to maximize the DRL computation efficiency and minimize the communication overhead (e.g., synchronization of model parameters). 
To achieve this goal, it requires achieving high-performance experience collection and model update simultaneously. While DRL serving presents an efficient template for experience collection, we still have to carefully consider the appropriate placement of trainers in sync. DRL training.

The first design option is to extend the \textit{task-dedicated GMIs} ($\mathit{TDG_{EX}}$) design strategy by including dedicated GMIs for DRL trainers. 
It will first gather experience from all serving GMIs (with an environment simulator and an agent) for trainer-side policy training. Once the policy training completes, all trainers will push the latest parameters for updating the policy model on agents. This design enables sharing trainer GMIs to serve multiple serving GMIs.
However, the latency-sensitive synchronized DRL training can hardly tolerate the high overhead of transferring experience across GMIs.
Besides, experience is not communication-friendly, since it consists of heterogeneous types of data (e.g., states, actions, and rewards) that are generated separately and vary in shapes.

A more promising direction is to align environment, agent, and trainer in the same GMI for fully GMI-based data-parallel training. This can be seen as the extended task-colocated GMI ($\mathit{TCG_{EX}}$). Different GMI would maintain a complete pipeline of all DRL training tasks, namely \textit{holistic training GMI}.
%
%
%
%
As shown in Figure~\ref{fig: GMI Mapping Template for Sync. and Async. DRL Training.}(a), 
a holistic training GMI introduces one more stage (iii) \textit{Global Policy Synchronization} to synchronize policy parameters in different GMIs. 
Once it completes, the local parameter of each GMI will be updated correspondingly. 
We also provide a comparison between $\mathit{TCG\_EX}$ and $\mathit{TDG\_EX}$. 
\begin{table}[t]
\centering
\caption{$\mathit{TDG_{EX}}$ and  $\mathit{TCG_{EX}}$ in synchronized DRL training.}
\vspace{-5pt}
\scalebox{0.85}{
\begin{tabular}{l|>{\centering\arraybackslash}p{4cm}|>{\centering\arraybackslash}p{3cm}}
\specialrule{.1em}{.05em}{.05em} 
\textbf{Solution} 
& \textbf{Resource Size} (\textbf{$R^{\mathbb{I}}$}) 
& \textbf{Comm. Size} (\textbf{$COM$}) 
\\ \hline\hline 
$\mathit{TDG_{EX}}$ 
& \small{$(T_s\cdot R_s+T_a\cdot\alpha\cdot R_a + T_t\cdot\beta\cdot R_t)/(T_s+T_a+T_t)$}
& $ m\cdot (S + A + W) + M_p$ + $ 2\cdot(n - 1)\cdot M_p/n$ 
\\ \hline 
$\mathit{TCG_{EX}}$ 
& \small{$(T_s+T_a+T_t)\cdot\max\{R_s, R_a, R_t\}/(T_s+T_a+T_t)$} 
& $2\cdot (n - 1)\cdot M_p/n$  \\ 
\specialrule{.1em}{.05em}{.05em} 
\end{tabular}}
\label{tbl: comparison between TDGex and TCGex in RL serving}
\vspace{5pt}
\end{table}
Our empirical studies show that $\alpha$ is around 0.2, $\beta$ is around 0.3, $R_s\approx10R_a\approx5R_t$,  and $T_s\approx6T_a\approx3T_t$, the overall system throughput is calculated as 
\begin{equation} \small \label{equ: top_train}
    \mathit{TOP_{train}} = \frac{R^{\mathbb{I}}_{all}}{R^{\mathbb{I}}}\cdot\frac{1}{T_s+T_a+T_t+\frac{COM}{BW}}
\end{equation}
Considering Table~\ref{tbl: comparison between TDGex and TCGex in RL serving} and $\frac{COM}{BW} \approx 7 \cdot (T_s+T_a+T_t)$ based on our profiling, we can derive that the overall system throughput of our $\mathit{TCG_{EX}}$ would increase evidently (about $5\times$) compared with $\mathit{TDG_{EX}}$. This is because the resource penalty on throughput 
($(T_s + T_a + T_t)\cdot\max\{R_s, R_a, R_t\}/(T_s\cdot R_s + T_a\cdot\alpha\cdot R_a +  T_t\cdot\beta\cdot R_t) - 1 \approx 0.5\times$) can be largely offset by significant throughput benefits (around $8\times$) via avoiding unnecessary inter-GMI experience movements. 

\textbf{Asynchronized DRL Training: } 
Asynchronized DRL training (e.g., A3C~\cite{a3c}) is another popular DRL training paradigm.
Compared to synchronized DRL training, asynchronized DRL concurrently executes DRL serving and policy training for experience collection and policy update.
In this case, the model update will no longer block the execution of policy serving. 
However, it comes with the increased staleness of policy parameters that may impact the algorithmic efficiency (e.g., convergence rate). 
The common design practice in asynchronized DRL training is to maximize the training throughput so that all DRL agents can get the updated model that can reflect the latest-collected experience.
Therefore, how to efficiently move experience between agent GMI and trainer GMI would be the key.
We introduce the decoupled serving and training scheme for GMI mapping (Figure~\ref{fig: GMI Mapping Template for Sync. and Async. DRL Training.}(b)), where serving GMI (for experience collection) are colocated on a set of GPUs to maximize the serving throughput, while the training GMIs are grouped on another set of GPUs to maximize training throughput. 
The solution to address the performance of experience collection and trainer synchronization has been detailed in the above DRL serving and $\S$\ref{sect: Layout-aware Gradient Reduction}.
Since asynchronous training has been offloaded to GMIs on other GPUs, the challenge now becomes how to efficiently move experience between serving and training GMIs ($\S$\ref{sect: Channel-based Experience Sharing}).

\subsection{Workload-aware GMI Selection} \label{sect: Workload-aware GMI Selection}
In addition to exploring highly efficient design templates, properly matching GMIs with DRL workloads is also of great importance and requires non-trivial efforts. 
Specifically, we need to (i) exploit the relation between GMI resource budget and DRL workload performance; (ii) balance the local GMI performance and the global system performance.
To this end, we introduce a \textit{workload-aware GMI selection} to facilitate the search of the right configuration of DRL simulation engine and the resource budget of GMI to achieve our key design objective (maximizing computation throughput). 

We consider two key factors that will impact the performance and resources of GMI-based DRL. 
i) \hlp{num\_env} describes the number of concurrent environments running at each GMI. It is equivalent to the batch size in conventional DNN training. The common choices for $num\_env$ range from 128 to 16,384 in consideration of the algorithmic (convergence) and system (runtime execution) efficiency~\cite{isaac-gym}. Generally, a larger value of $num\_env$ would lead to higher computation throughput. 
However, when all the computation resources of a GMI are fully occupied, further increasing the $num\_env$ would only increase the memory consumption while being largely throttled  by computation resources. This would lead to minor or even no throughput performance improvement.
ii) \hlp{GMIperGPU} describes the quota of computation resources assigned to each GMI. The size of computation resources would determine whether GMI training can be initiated and executed successfully, and more importantly, the throughput performance of DRL computation. 
This is because the simulation and the training process are computation intensive. Memory resources, on the other hand, could only determine whether the DRL training can be accommodated on GPU.
\begin{algorithm}[t] \footnotesize
\setstretch{0.9} 
  \caption{Profiling-based GMI Exploration.}
  \label{algo: Profiling-based GMI Exploration.}
\SetAlgoLined
  \SetKwInOut{Input}{input}
  \SetKwInOut{Output}{output}
  \Input{$\mathit{DRL\_bench}$, $\mathit{num\_GPU}$}
  \Output{$\mathit{num\_env}$, $\mathit{GMIperGPU}$}
    $\mathit{best\_config} = \mathit{tuple}()$;  
    $\mathit{max\_top} = -\mathit{inf}$\;
    \For{$\mathit{GMIperGPU}$ \textbf{in} 10 ... 1}{
        $\mathit{pre\_top}=0$; $\mathit{pre\_mem}=0$\;
        \For{$\mathit{num\_env}$ \textbf{in} [128, 256, 512, ..., 16384]}  { 
        \tcc{Profile performance of a GMI.}
        $\mathit{runnable}$, $\mathit{top}$, $\mathit{mem}$ = \textbf{profile}($\mathit{DRL\_bench}$, $\mathit{GMIperGPU}$, $\mathit{num\_env}$)\;
        \tcc{Filter out non-runnable GMI.}
        \If{\textbf{not} $\mathit{runnable}$}{
            \textbf{continue}\;
        }
        \tcc{Initialize tracking variables.}
        \If{$\mathit{pre\_top}==\mathit{pre\_mem}==0$}{
            $\mathit{pre\_top} = \mathit{top}$; 
            $\mathit{pre\_mem} = \mathit{mem}$\; 
            \textbf{continue}\;
        }
        \tcc{Compute performance/resource changes.}
        $\mathit{R_{top}} = (\mathit{top} - \mathit{pre\_top})/(\mathit{pre\_top})$\;
        $\mathit{R_{mem}} = (\mathit{mem} - \mathit{pre\_mem})/(\mathit{pre\_mem})$\;
        $\mathit{Sat} = \mathit{R_{top}} / \mathit{R_{mem}}$\;
        $\mathit{pre\_top} = \mathit{top}$; $\mathit{pre\_mem} = \mathit{mem}$\;
        \tcc{Check if the performance saturates.}
        \If{$\mathit{Sat} < \alpha$}{
            \textbf{break}\;
        }
        \tcc{Project the overall system throughput.}
        $\mathit{acc\_top}$ = \textbf{estimate}($\mathit{GMIperGPU}$, $\mathit{num\_GPU}$, $\mathit{top}$)\;
        \If{$\mathit{acc\_top}$ > $\mathit{max\_top}$}{
             $\mathit{max\_top} = \mathit{acc\_top}$\;
             $\mathit{best\_config} = (\mathit{num\_env}, \mathit{GMIperGPU})$\;
        }
       }
    }
    $\mathit{num\_env}$, $\mathit{GMIperGPU}$ = $\mathit{best\_config}[0]$, $\mathit{best\_config}[1]$\; 
    \textbf{return} $\mathit{num\_env}$, $\mathit{GMIperGPU}$;
\end{algorithm}

Our approach will take the DRL environment specification (e.g., name and type) and the number of GPUs as the input, and return $\mathit{num\_env}$ and $\textit{GMIperGPU}$ as desired runtime configuration.
As illustrated in Algorithm~\ref{algo: Profiling-based GMI Exploration.}, we will iterate through the design space of GMI resource sizes by adjusting the number of GMIs per GPU (Line 2). For instance, when $\mathit{GMIperGPU}=4$, the resources of each GPU will be divided into 4 GMIs evenly. Another design dimension to be explored is the number of environments ($\mathit{num\_env}$). This will help us to pinpoint the design that can maximize the throughput performance given the GMI resource budget. To reduce unnecessary search, we leverage several metrics to prune out design points and use heuristic rules for an early stop.

The \textit{profile} function will test out whether the DRL benchmark is runnable under the current GMI resource and $\mathit{num\_env}$ configuration. 
If it fails to run (e.g., hanging and crashing), which indicates insufficient computation/memory resources, we will skip all the following steps for performance analysis in the current step. 
Otherwise, if it is runnable, we will compare the change of computation throughput and the change of memory consumption between its current successful run and its last successful run (Line 13 to 14). 
To determine whether GMI has already reached its maximum capability for DRL simulation, we introduce a saturation ($\mathit{Sat}$) metric by normalizing the \textit{throughput gains} with respect to the \textit{memory consumption penalty} (Line 15). 
Our empirical study ($\S$\ref{sect: additional studies} for concurrent environments study) shows that a GMI tends to saturate its computation capacity when we increase the $\mathit{num\_env}$ and observe (i) the large increase of memory consumption (ii) the minor improvement of computation throughput.  
We use such insights to set an early stop signal by comparing the $\mathit{Sat}$ against a threshold value $\alpha$ (generally $\alpha<0.1$). If the current GMI computation capacity is yet to be saturated ($\mathit{Sat}\geq\alpha$), we will estimate the overall system throughput and record the best setting if the estimated throughput is larger than any previously searched designs (Line 20 to 24). 
\vspace{-5pt}
\section{Evaluation}
\begin{table}[t] \small
\caption{DRL Benchmarks and Policy NN Models.}
\vspace{-5pt}
\centering
\scalebox{0.89}{
\begin{tabular}{l|c | c |  r | l}
\specialrule{.1em}{.05em}{.05em} 
\textbf{Benchmark} &\textbf{Abbr.} & \textbf{Type} & \textbf{\#Dim.} & \textbf{Policy NN Model} \\ 
\hline
Ant             & AT & L& 60  & 60:256:128:64:8    \\ 
Anymal          & AY & L& 48 & 48:256:128:64:12 \\
BallBalance     & BB & L& 24 & 24:256:128:64:3 \\
FrankaCabinet   & FC & F & 23 & 23:256:128:64:9 \\
Humanoid        & HM & L& 108 & 108:200:400:100:21 \\
ShadowHand      & SH & R& 211 & 211:512:512:512:256:20 \\
\specialrule{.1em}{.05em}{.05em} 
\end{tabular}}
\label{tbl: RL Benchmark Settings.}
\vspace{2pt}
\end{table}

\begin{figure*}[t] \small
    \centering
    \vspace{-5pt}
    \subfloat[]{\includegraphics[width=0.33\textwidth]{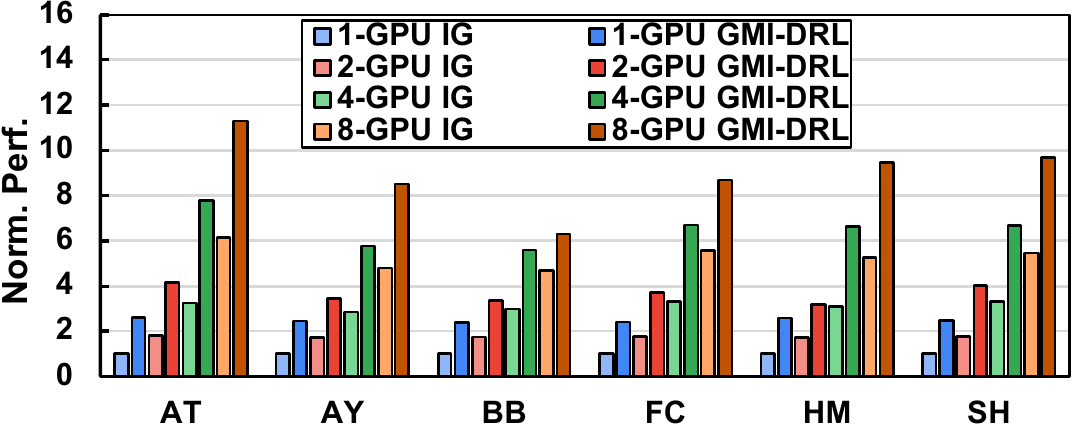}}
    \subfloat[]{\includegraphics[width=0.33\textwidth, trim=0 0cm 0 0]{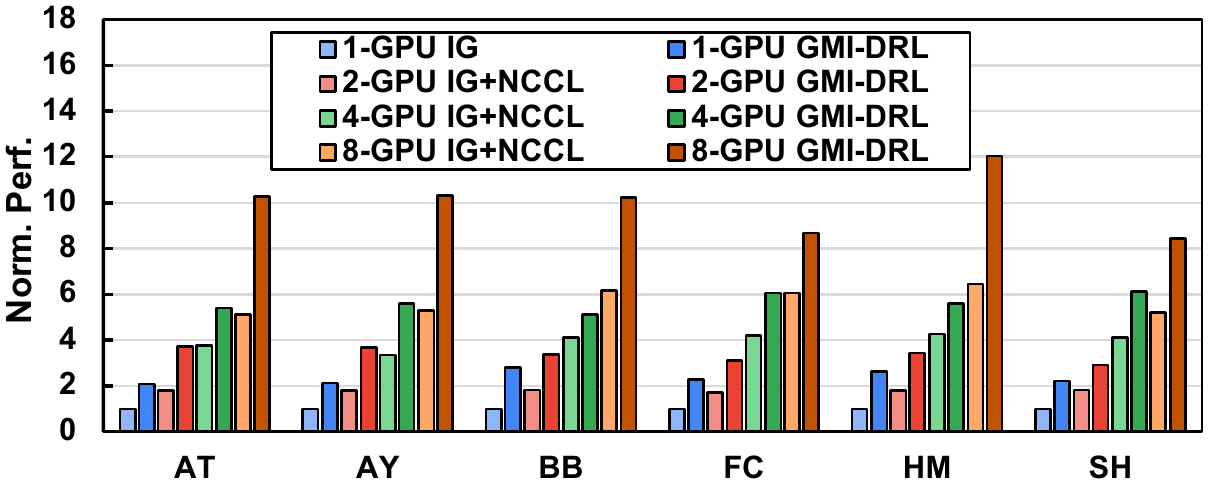}}
    \subfloat[]{\includegraphics[width=0.33\textwidth,  trim=0.1cm 0cm 0 0]{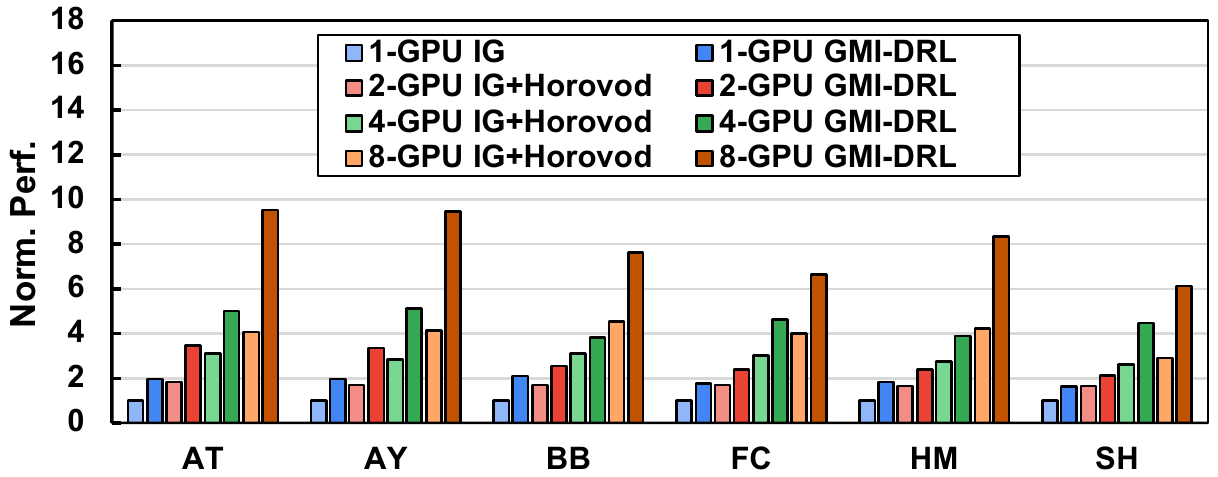}}
    \vspace{-10pt}
    \caption{End-to-end computation throughput comparison for (a) \textit{DRL serving}, (b) \textit{Sync. DRL training with NCCL}, and (c) \textit{Sync. DRL training with Horovod}. Note that computation throughput is normalized w.r.t. NVIDIA Isaac Gym on a single GPU.}
    \label{fig: End-to-End performance comparison}
    \vspace{-10pt}
\end{figure*}
\textbf{Benchmarks: } The DRL benchmark consists of two major components: the \textit{simulation environment} and the \textit{policy NN model}. 
We cover different types of environments, including locomotion simulation~\cite{liang2018gpu,hwangbo2019learning} (\textbf{L}), franka cube stacking~\cite{khatib1987unified} (\textbf{F}), robotics hand control~\cite{andrychowicz2020learning} (\textbf{R}).
%
The details of our evaluated DRL benchmarks are listed in Table~\ref{tbl: RL Benchmark Settings.}. 
``\#Dim.'' is the dimension of environment state vectors. 
The policy NN models are expressed in the format of ``\textit{in\_dim:hidden\_dim:...:out\_dim}''. 
In this experiment, we will mainly focus on the synchronized DRL training (PPO~\cite{ppo} algorithm), which is the official implementation from Isaac Gym~\cite{isaac-gym}. 
We also leverage Isaac Gym to implement A3C~\cite{ga3c} for an async. DRL training study.

\textbf{Implementations: } 
We use several baselines for evaluation. 
\textit{\textbf{(i) Isaac Gym multi-GPU serving}} is to directly scale the Isaac Gym serving on multi-GPU platform where each GPU is occupied by one Isaac Gym serving process; 
\textit{\textbf{(ii) Isaac Gym (PPO) + NCCL}} is for synchronized DRL training in data-parallel fashion, and uses NCCL~\cite{nccl} as the communication backend;
\textit{\textbf{(iii) Isaac Gym (PPO) + Horovod}} is also for synchronized DRL training in data-parallel fashion, and uses Horovod~\cite{sergeev2018horovod} as the communication backend;
\textit{\textbf{(iii) Isaac Gym (A3C)}} is for study on asynchronized DRL training, where policy serving (experience collection) and policy training are running concurrently on different GPUs.
Note that for all baselines, we manually adjust the simulation batch size (i.e., the number of concurrent environments) of each DRL benchmark to reach its peak throughput by using each GPU exclusively (i.e., no sharing with other processes).
The runtime configurations (e.g., $\mathit{num\_env}$ and $\mathit{GMIperGPU}$) of our \Mname~implementation are generated by using Algorithm~\ref{algo: Profiling-based GMI Exploration.}.

\textbf{Platforms \& Tools: } 
We use a DGX-A100~\cite{dgx-a100} (including 8$\times$NVIDIA A100 GPUs and Dual AMD Rome 7742 CPUs with 128 cores@2.25 GHz base clock). For environment simulation, we leverage NVIDIA Isaac Gym~\cite{isaac-gym}. We use PyTorch (v3.8)~\cite{pytorch} to build the policy model, and NCCL (v2.8.4)~\cite{nccl} and Gloo~\cite{gloo} (the latest commit on \texttt{main} branch) to build the GMI-based communication layer. 
To measure the performance, we leverage \texttt{nvidia-smi} tool and PyTorch runtime profiler. The throughput metric of DRL serving and sync. DRL training is the simulation steps per second~\cite{isaac-gym}.

\vspace{-5pt}
\subsection{Overall Performance} 
\label{sect: Overall Performance}

\textbf{DRL Serving } 
DRL serving is important for accumulating experiences for offline policy training and finetuning, especially for those DRL tasks that online policy training would compromise user experience and safety, such as autonomous driving.
As shown in Figure~\ref{fig: End-to-End performance comparison}(a), with the help of our GMI-oriented design, the overall throughput of policy serving can be improved (up to $2.62\times$, $2.08\times$ on average) across settings with different numbers of GPUs. Based on our detailed profiling results, our GMI-based design can improve the GPU utilization (up to 45.7\%, 27.9\% on average) compared with the state-of-the-art Isaac Gym. This also helps to demonstrate the benefits of using our \Mname~framework to maximize GPU utilization and performance gains. 

\textbf{Sync. DRL Training} 
To measure the performance of the synchronized DRL training, we follow the convention of using steps per second as the performance metric~\cite{isaac-gym}. 
%
%
\hlp{Compared with Isaac Gym (PPO) + NCCL:}  Figure~\ref{fig: End-to-End performance comparison}(b) demonstrates that \Mname~with the optimal GMI planning (selection, customization, and mapping) can achieve significant improvements (up to $2.81\times$, $1.86\times$ on average) compared to Isaac Gym with NCCL under the same number of GPUs. 
We also observe that in the more complex DRL setting with a higher number of concurrent simulations and state/observation dimensions, our GMI-based design could demonstrate more benefits.
This is because of the effective 1) resource planning to harvest the performance gains from available GPU resources, where \Mname~improves $31.8\%$ GPU utilization on average; 2) resource isolation to avoid execution interference. Moreover, \Mname~can help users to save more than 50\% cloud service cost on average comparing the baseline approach without GMI-based design when achieving the same training throughput.
\hlp{Compared with Isaac Gym (PPO) + Horovod:} Horovod~\cite{sergeev2018horovod} is the state-of-the-art communication library for multi-GPU model training. 
%
%
Figure~\ref{fig: End-to-End performance comparison}(c) shows that \Mname~outperforms Isaac Gym with Horovod backend with up to $2.34\times$ (1.75$\times$ on average) training performance throughput.
This result implies that only focusing on optimizing communication is insufficient to exploit the full potential of multi-GPU platforms. In contrast, \Mname~systematically addresses the GPU underutilization and communication bottleneck by introducing GMI-based design and GMI-tailored communication support. Thus, \Mname~achieves evidently higher performance than Horovod baseline.

\begin{figure} [t] \small
    \centering
    \includegraphics[width=\columnwidth]{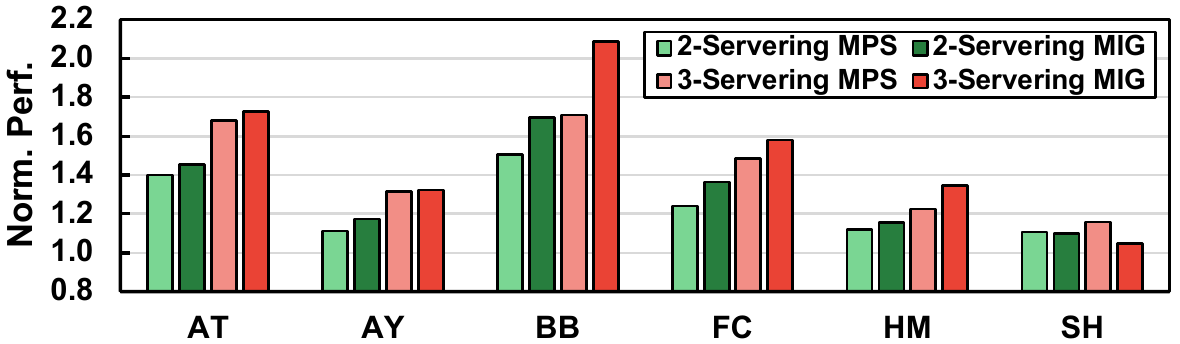}
    \vspace{-15pt}
    \caption{GMI backend comparison between MPS and MIG. Note that the performance of MPS and MIG is normalized \textit{w.r.t.} direct GPU sharing among processes.}
    \label{fig: backend-choice}
    \vspace{5pt}
\end{figure}

\vspace{-5pt}
\subsection{Additional Studies}
\label{sect: additional studies}
\textbf{GMI Backend} 
We will demonstrate the different backend choices: \textit{Direct-Share} (direct GPU sharing among processes), \textit{MPS}, and \textit{MIG}. 
We select the setting of 2-Serving and 3-Serving on 1$\times$A100 for study.
Figure~\ref{fig: backend-choice} shows that \textit{MPS} and \textit{MIG} consistently outperform \textit{Direct-Share}. 
The major reasons behind this are that (i) environment simulation consists of intensive and resource-demanding computation and (ii) \textit{MPS} and \textit{MIG} can offer better resource isolation to reduce memory and computation contention among processes.
On more complicated DRL benchmarks (e.g., HM and BB), the hardware-level isolation of \textit{MIG} would outperform \textit{MPS} with server-client-based resource multiplexing because of the better service-level performance guarantee of \textit{MIG}. 
Whereas on the simpler DRL benchmarks (e.g., AT), the performance difference is minor between \textit{MPS} and \textit{MIG}, since the lightweight workload is less likely to cause GPU resource contention.
\begin{figure*}[t] \small
    \centering
    \subfloat{\includegraphics[width=0.33\textwidth, height=2.5cm]{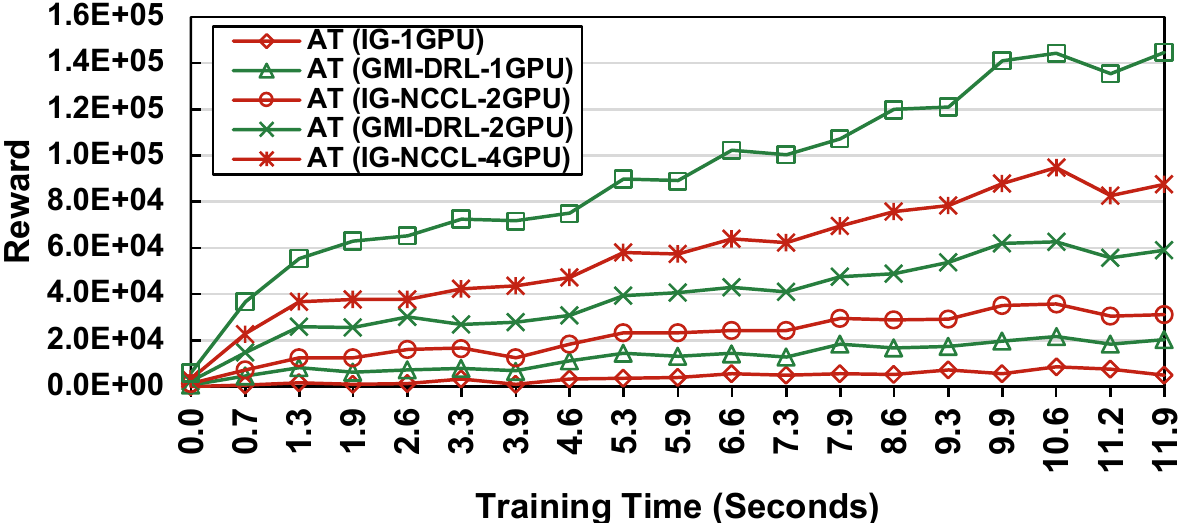}}
    \subfloat{\includegraphics[width=0.33\textwidth, height=2.5cm, trim=0 0cm 0 0]{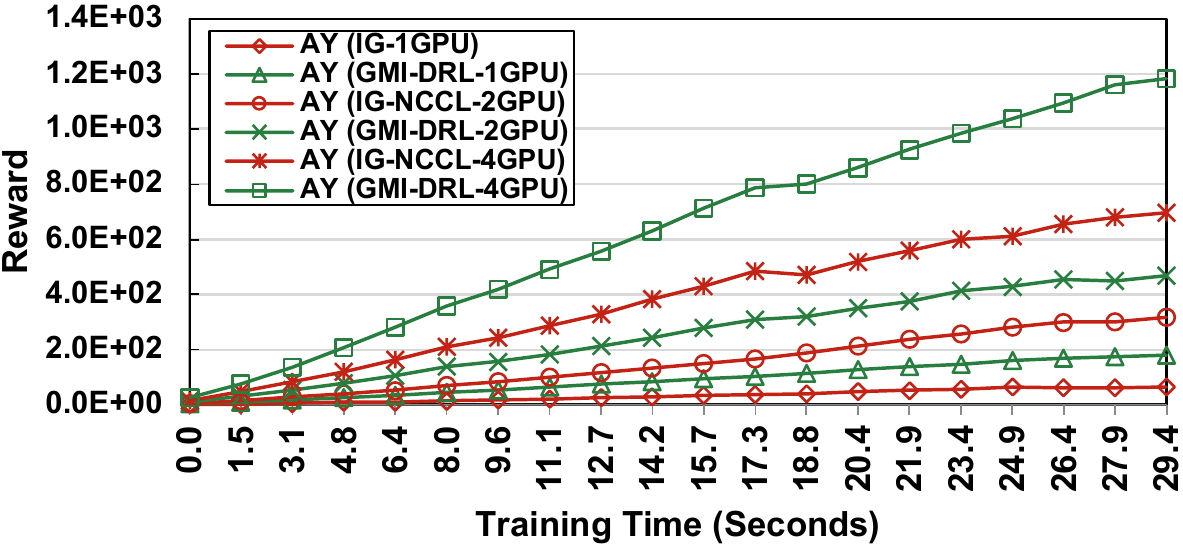}}
    \subfloat{\includegraphics[width=0.33\textwidth, height=2.5cm, trim=0 0.1cm 0 0]{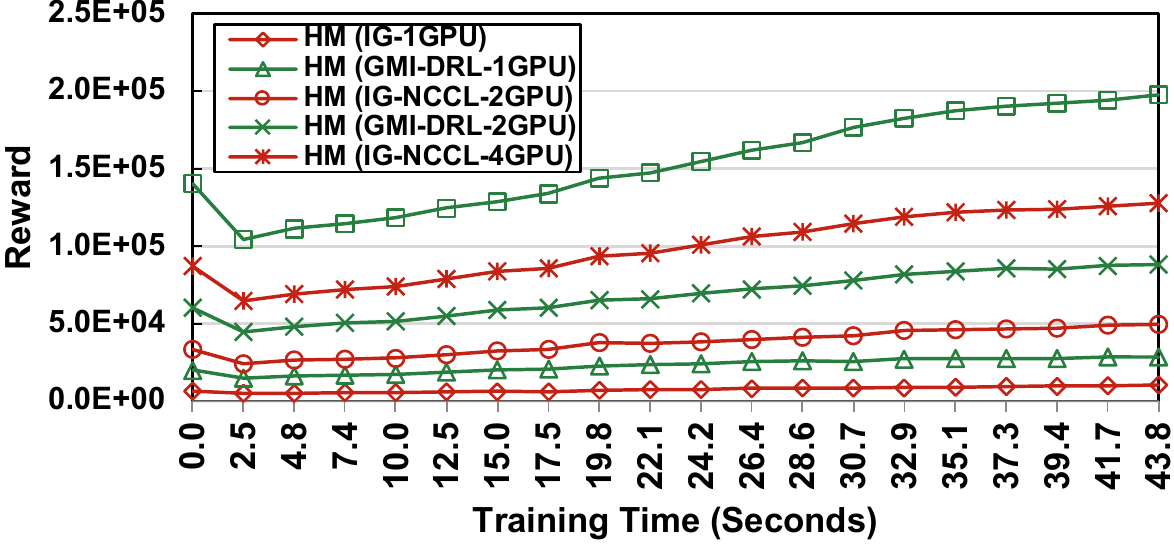}}
    \vspace{-5pt}
    \caption{Reward accumulation over training time (in seconds) for 20 epochs on {AT} (\textit{\textbf{left}}), {AY} (\textbf{\textit{middle}}), and {HM} (\textit{\textbf{right}}).}
    \label{fig: reward accumulation over training iterations}
    \vspace{-10pt}
\end{figure*}

\textbf{Benefits of LGR} We will show the effectiveness of our GMI-tailored layer-aware gradient reduction (LGR) on complex GMI layouts requiring cross-GPU communication. 
Our baseline is the design with MPR only. We choose the settings: 2 GPUs with 2 Trainers per GPU (2G2T), 2 GPUs with 3 Trainers per GPU (2G3T), and 4 GPUs with 4 Trainers per GPU (4G4T) for illustration and the number of DRL trainers is the same for both baseline and LGR design. 
Table~\ref{tbl: benefits of HGR.} indicates that LGR can bring throughput benefits for DRL training on models with different parameter sizes. We also notice the larger performance benefit under the more number of GPUs, since LGR can well parallelize the inter-GMI reduction and utilize high-bandwidth GPU interconnections (e.g., NVLink) for better communication performance. 
\begin{table}[t] \small
\centering
\caption{Comparing the throughput of \textit{LGR} and  \textit{MPR} baselines on Sync. DRL training.}
\vspace{-7pt}
\scalebox{0.7}{
\begin{tabular}{l|l|r|r|r|r|r|r}
\specialrule{.1em}{.05em}{.05em} 
\multirow{2}{*}{\textbf{Bench.}} & \multicolumn{1}{l|}{\multirow{2}{*}{\textbf{Param.}}} & \multicolumn{2}{c|}{\textbf{2G2T}} & \multicolumn{2}{c|}{\textbf{2G3T}} & \multicolumn{2}{c}{\textbf{4G4T}} \\ \cline{3-8} 
 & \multicolumn{1}{l|}{} & \multicolumn{1}{l|}{\textbf{Baseline}} & \multicolumn{1}{l|}{\textbf{LGR}} & \multicolumn{1}{l|}{\textbf{Baseline}} & \multicolumn{1}{l|}{\textbf{LGR}} & \multicolumn{1}{l|}{\textbf{Baseline}} & \multicolumn{1}{l}{\textbf{LGR}} \\ \hline\hline
\textbf{AT} & 1.1$\times10^5$ & \multicolumn{1}{r|}{107,689} & \textbf{114,734} & \multicolumn{1}{r|}{138,369} & \textbf{164,655} & \multicolumn{1}{r|}{168,619} & \textbf{207,834} \\ \hline
\textbf{HM} & 2.9$\times10^5$ & \multicolumn{1}{r|}{163,723} & \textbf{168,300} & \multicolumn{1}{r|}{179,273} & \textbf{190,919} & \multicolumn{1}{r|}{308,873} & \textbf{336,591} \\ \hline
\textbf{SH} & 1.5$\times10^6$ & \multicolumn{1}{r|}{78,270} & \textbf{88,149} & \multicolumn{1}{r|}{79,545} & \textbf{93,785} & \multicolumn{1}{r|}{133,044} & \textbf{166,722} \\ 
\specialrule{.1em}{.05em}{.05em} 
\end{tabular}}
    \label{tbl: benefits of HGR.}
    \vspace{-15pt}
\end{table}
\begin{figure}[t] \small
    \centering
    \subfloat{\includegraphics[width=0.245\textwidth]{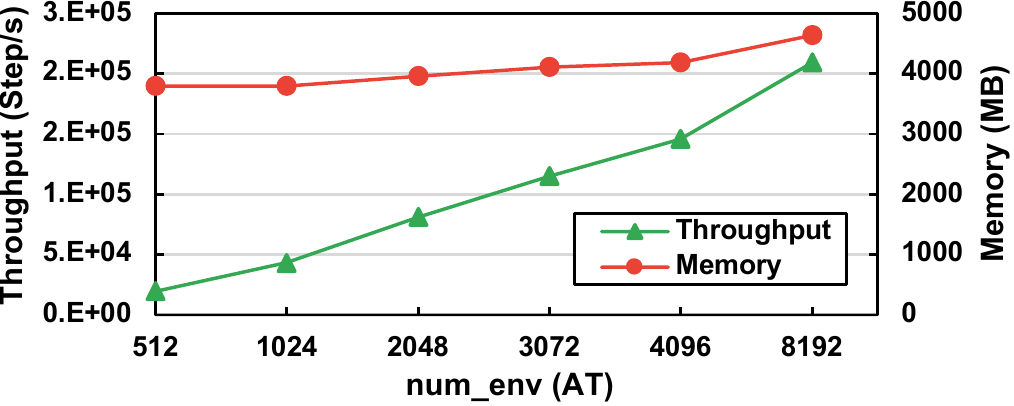}}
    \subfloat{\includegraphics[width=0.245\textwidth, trim=0 0cm 0 0]{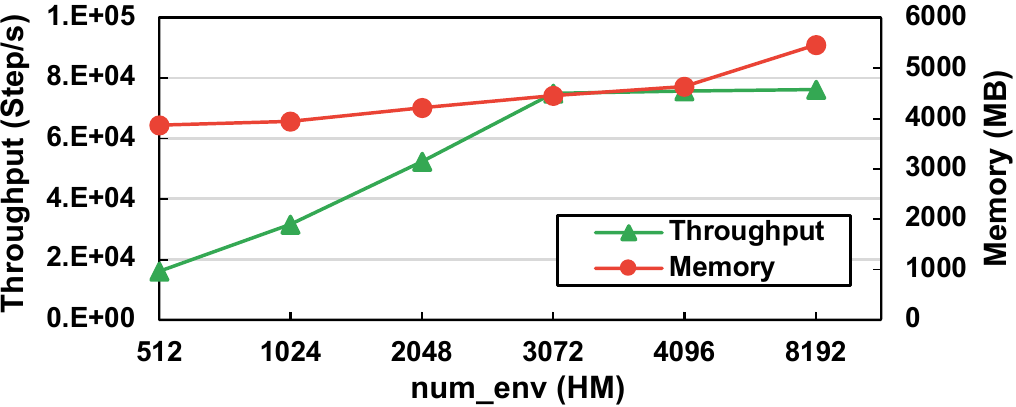}}
    \vspace{-5pt}
    \caption{Sync. DRL Training \textit{throughput} and GPU \textit{memory} w.r.t. \textit{num\_env} on AT (\textbf{\textit{left}}) and HM (\textbf{\textit{right}}).}
    \label{fig: Serving througput and memory w.r.t. num_envs.}
    \vspace{5pt}
\end{figure}

\begin{figure} [t] \small
    \centering
    \includegraphics[width=\columnwidth]{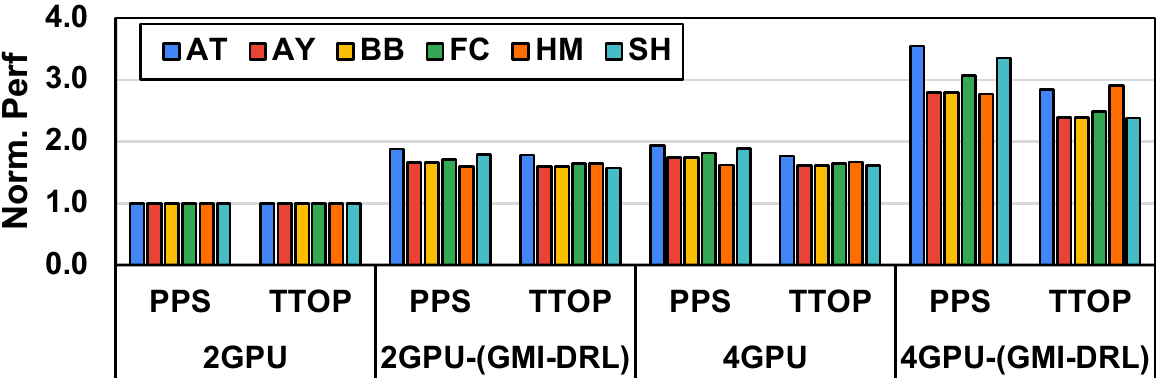}
    \vspace{-15pt}
    \caption{Async. DRL training throughput performance.}
    \label{fig: Async DRL training.}
    \vspace{5pt}
\end{figure}

\textbf{Concurrent Environment} 
We analyze the impact of the number of environments ($num\_env$) per simulator on sync. DRL training. We take AT and HM for illustration and use 1 GMI on 1 GPU to model its impact on throughput performance and memory consumption. We cover the common value range of $num\_env$ (from 512 to 8,192).
Table~\ref{fig: Serving througput and memory w.r.t. num_envs.} shows that the increasing number of environments will improve the training performance at the cost of larger memory consumption. 
We also notice that a larger \textit{num\_env} does not necessarily bring an equivalent amount of throughput improvement.
For instance, when \textit{num\_env} increases from 4,096 to 8,192, the throughput increases slightly while the GPU memory consumption increases sharply. Such an observation also agrees with our insights for workload-aware GMI searching ($\S$\ref{sect: Workload-aware GMI Selection}).

\textbf{Reward Accumulation} 
For training efficiency, we measure the reward accumulation speed over training time, following the common practice adopted by existing DRL work~\cite{isaac-gym, ga3c}. 
We choose three DRL benchmarks (AT, AY, and HM), which can cover representative DRL applications for local motion and robotics simulation.
Figure~\ref{fig: reward accumulation over training iterations} shows that within a given amount of training time, \Mname~achieves an evidently higher accumulated reward compared with the baseline single-GPU Isaac Gym, and its multi-GPU variants with NCCL backend. And the reward accumulation benefit of using \Mname~would also scale up with the increasing number of GPUs.
By collecting more rewards given the same GPU resource budget and training iteration constraints, the well-trained DRL policy can achieve much better performance (i.e., making high-reward decisions) in real-world settings.
\begin{table}[t] \small
\centering
\vspace{-12pt}
\caption{Comparing the throughput of \textit{Uni-channel} and \textit{Multi-channel} experience sharing.}
\vspace{-7pt}
\scalebox{0.8}{
\begin{tabular}{l|r | r | r | r}
\specialrule{.1em}{.05em}{.05em} 
\textbf{2 GPUs} & 
\textit{UCC$_{PPS}$}
& \textit{MCC$_{PPS}$}  
& \textit{UCC$_{TTOP}$}
& \textit{MCC$_{TTOP}$} 
\\ \hline\hline
AY & \multicolumn{1}{r|}{169,451} & \multicolumn{1}{r|}{\textbf{180,001}} & \multicolumn{1}{r|}{108,536} & \textbf{122,676} \\ \hline
FC & \multicolumn{1}{r|}{98,819} & \multicolumn{1}{r|}{\textbf{102,363}} & \multicolumn{1}{r|}{40,778} & \textbf{50,931} \\ 
\specialrule{.1em}{.05em}{.05em} 
\textbf{4 GPUs} & 
\textit{UCC$_{PPS}$}
& \textit{MCC$_{PPS}$}  
& \textit{UCC$_{TTOP}$}
& \textit{MCC$_{TTOP}$} 
\\ \hline\hline
AY & \multicolumn{1}{r|}{287,990} & \multicolumn{1}{r|}{\textbf{312,366}} & \multicolumn{1}{r|}{174,569} & \textbf{215,876} \\ \hline
FC & \multicolumn{1}{r|}{171,015} & \multicolumn{1}{r|}{\textbf{182,010}} & \multicolumn{1}{r|}{67,027} & \textbf{88,622} \\ 
\specialrule{.1em}{.05em}{.05em} 
\end{tabular}}
\label{tbl: multi-channel communication}
\vspace{5pt}
\end{table}

\textbf{Async. DRL Training} 
We compare \Mname~with non-GMI solutions in asynchronized DRL training. 
We measure two major metrics: the \textit{predictions per second} (PPS) and the \textit{training samples throughput} (TTOP). 
Figure~\ref{fig: Async DRL training.} shows that with the help of our GMI-based design and the throughput optimized experience transferring technique, \Mname~can achieve averaged $1.88\times$ PPS and $1.65\times$ (TTOP) improvement over non-GMI baseline across 2-GPU and 4-GPU settings. 
This result demonstrates the importance of collaboratively improving the performance of experience collection and experience sharing in accelerating async. DRL training.

\textbf{Multi-Channel Experience Sharing}
We measure the effectiveness of our multi-channel design (MCC) by comparing it with the uni-channel solution (UCC). 
We choose AY and FC on 2 GPUs and 4 GPUs for this study. 
And we evaluate both the PPS and TTOP for both implementations. Table~\ref{tbl: multi-channel communication} shows that our MCC solution can improve PPS and TTOP evidently compared with UCC in A3C training.
This can be attributed to the improved memory bandwidth utilization, since UCC-based experience sharing would incur lots of fine-grained data communication so that memory bandwidth is largely underutilized. In contrast, our MCC solution increases the data movement granularity and cross-GMI bandwidth utilization through effective data categorization and batching.
\section{Related Work}
\vspace{-5pt}
The surge of deep reinforcement learning has raised the interest of system researchers and engineers to improve its performance. 
In one direction, an array of CPU-based distributed frameworks have been developed for DRL training.
Acme~\cite{hoffman2020acme} is a CPU-based RL research framework for running RL algorithms at different scales. Seed RL~\cite{espeholt2019seed} is a distributed RL framework with batched inference/training by leveraging CPU-based simulator and TPU-based agent/trainer.
Ray~\cite{ray} is a distributed computing framework encompassing a dedicated RL package, RLlib~\cite{liang2017rllib}. 
RLlib uses the CPU-based simulator in collaboration with Tensorflow/PyTorch as the backend for the NN operations of agents and trainers. 
The DRL performance on Ray is largely throttled by the performance of its CPU simulator, which depends on CPU threads/processes for parallelization. We also explore the possibility of using GPU-based simulators~\cite{cule, isaac-gym} in Ray, but it always crashes the GPU simulator due to inappropriate scheduling of the GPU-based simulation tasks. In short, these designs are largely hindered by their slow CPU-based simulators.

Another direction leverages the power of GPU to accelerate DRL.
GA3C~\cite{ga3c} proposes a CPU-GPU training architecture for the A3C algorithm~\cite{a3c} by accelerating the action prediction and policy training on GPUs while keeping the environment simulation on CPUs. 
cuLE~\cite{cule} spots environment simulation as the major performance bottleneck of DRL systems and implements a GPU-based RL environment for fast simulation. 
NVIDIA Flex~\cite{liang2018gpu} introduces a GPU-based simulator and distributed RL simulation and training via direct data parallel.
Isaac Gym~\cite{isaac-gym} represents NVIDIA's latest single-GPU-based design for massively parallel RL environment simulation. 
Isaac Gym achieves state-of-the-art performance (200$\times$ faster) over those CPU-based designs for DRL training.
Even though the environment simulation has already been accelerated by GPUs, the performance of environment simulation is still the key performance bottleneck (Figure~\ref{fig: Performance profiling of isaac-gym on GPU utilization during the training.}). 
This is because the throughput performance of a single GPU-based simulation engine (with the complex physical calculation) cannot be improved as well as the regular dense-linear algebraic computation (with GEMM-based computation).

\vspace{-5pt}
\section{Discussions}
\vspace{-5pt}
Our work would potentially benefit several future directions:

\hlp{For potential users}, our work provides good practice to extend GPU spatial multiplexing technique to reach a broader range of applications:
1) characterizing the workload of GPU tasks through lightweight micro-benchmarking. The key for this step is to model the relation between the task performance and GPU resource cost; 
2) matching tasks with right-fit GPU resources through a tradeoff between resource consumption and performance; 
3) co-locating communication-intensive tasks to GMIs to minimize the cost of data movement. 

\hlp{For GPU vendors}, we envision the great potential of the more efficient \textit{inter-process intra-GPU} communication technique (such as dedicated device memory space shared among GMIs) to further reduce the cost of data communication among GMIs. 
Besides, \textit{dynamically-adjustable instance size} will be another key feature for adapting GPU instances to accommodate some special applications, which have different computation and memory resource demands at different time frames during the runtime execution.

\hlp{For DRL scaling}, our work can be a good starting point for \textit{scaling up} and \textit{scaling out} DRL training. 
For scaling up on a single multi-GPU node, our adaptive GMI management strategy can guide the design of tasks-to-GMI mapping and GMI-to-GPU layout to maximize the intra-node throughput. 
For scaling out to multiple multi-GPU nodes, our layout-aware gradient reduction technique can be extended to support efficient multi-node model synchronization by considering the intra- and inter-node GMI layout hierarchy.
Our multi-channel communication can be optimized for diverse types of data paths (e.g., NVlink and InfiniBand) so that we can maximize communication efficiency and bandwidth utilization.

\hlp{For cluster scheduling}, our work (focusing on one GPU job) would also improve the efficiency of the existing cluster schedulers~\cite{xiao2018gandiva, xiao2020antman} for handling multiple independent GPU jobs. 
One key challenge of the existing cluster scheduling design is the non-adjustable GPU computation layout (e.g., low GPU utilization per job) that introduces resource underutilization and sub-optimal system performance. 
\Mname~would open up new design dimensions for future cluster schedulers to optimize their \textit{GPU computation layout} (e.g., condensing fragmented GPU jobs into fewer GPUs) and \textit{dynamic GPU computation adjustment} (e.g., recycling and reshaping spare GPU resources to fit jobs with the GPU-affinity demand). This will reduce the idleness of GPU resources and improve the cluster scheduling flexibility.

\section{Conclusion}
In this paper, we propose GMI-DRL, a systematic design to accelerate DRL on GPUs with instance-based GPU spatial multiplexing. It includes a novel design of resource-efficient GPU multiplexing instances (GMIs), a process-based GMI programming scheme, and an adaptive GMI management strategy for high GPU utilization and system throughput. We also build highly efficient communication support across GMIs to meet various communication demands of DRL training.
Comprehensive experiments demonstrate the \Mname~advantages 
over the state-of-the-art Isaac Gym on a single GPU and its variants with NCCL/Horovod on multi-GPU settings.

\bibliographystyle{plain}
\bibliography{reference}

\end{document}